\documentclass[showpacs,twocolumn,aps,preprintnumbers,letterpaper]{revtex4}
\usepackage{amsmath,amssymb}
\usepackage{epsfig}
\usepackage{graphicx}
\usepackage{amsmath}
\usepackage{slashed}
\usepackage{amsfonts}
\usepackage{epstopdf}

\newcommand{\be}{\begin{equation}}
\newcommand{\ee}{\end{equation}}
\newcommand{\bear}{\begin{eqnarray}}
\newcommand{\eear}{\end{eqnarray}}
\newcommand{\ba}{\begin{array}}
\newcommand{\ea}{\end{array}}

\def\be{\begin{eqnarray}}
\def\ee{\end{eqnarray}}
\def\bea{\be}
\def\eea{\ee}

\newcommand{\e}{{\mbox{e}}}

\def\roughly#1{\mathrel{\raise.3ex\hbox{$#1$\kern-.75em%
\lower1ex\hbox{$\sim$}}}}

\begin{document}

\title{Light Quarks in the  Screened Dyon-Anti-Dyon  Coulomb Liquid Model II}

\author{Yizhuang Liu}
\email{yizhuang.liu@stonybrook.edu}
\affiliation{Department of Physics and Astronomy, Stony Brook University, Stony Brook, New York 11794-3800, USA}

\author{Edward Shuryak}
\email{edward.shuryak@stonybrook.edu}
\affiliation{Department of Physics and Astronomy, Stony Brook University, Stony Brook, New York 11794-3800, USA}

\author{Ismail Zahed}
\email{ismail.zahed@stonybrook.edu}
\affiliation{Department of Physics and Astronomy, Stony Brook University, Stony Brook, New York 11794-3800, USA}


\date{\today}
\begin{abstract}
We discuss an extension of the dyon-anti-dyon liquid model that includes light quarks in the  dense center symmetric
 Coulomb phase. In this work, like in our previous one, we use the simplest
 color SU(2) group. We start with a single fermion flavor $N_f=1$ and
explicitly map the theory onto a 3-dimensional quantum effective theory with a fermion that is only
U$_V(1)$ symmetric. We use it to show that the dense center symmetric plasma
develops, in the mean field approximation, a nonzero chiral condensate, although the ensuing Goldstone mode is massive due to the U$_A(1)$ axial-anomaly.
 We estimate  the chiral condensate and $\sigma,\eta$  meson
 masses for $N_f=1$. We then extend our analysis to several flavors $N_f>1$ and 
 colors $N_c>2$  and show that center symmetry and   spontaneous chiral symmetry breaking  disappear simultaneously when $x=N_f/N_c\geq 2$
 in the dense plasma phase.  A reorganization of the dense plasma phase into a
 gas of dyon-antidyon molecules restores  chiral
 symmetry, but may preserve center symmetry in the linearized approximation. We estimate  the corresponding critical temperature.

\end{abstract}


\maketitle

\setcounter{footnote}{0}


\section{Introduction}

This work is a continuation of our earlier study~\cite{LIU1}  of the  gauge 
topology in the confining phase of a theory with the simplest gauge group $SU(2)$.  
We suggested that the confining phase below the transition temperature is an
   ``instanton dyon" (and anti-dyon) plasma which is dense enough
to generate strong screening. The dense plasma is amenable to 
standard mean field methods.

While an extensive introduction to the subject can be found in~\cite{LIU1}, here we only mention few
important points. The treatment of the gauge topology near and below $T_c$ is based on the discovery of KvBLL
instantons threaded by finite holonomies~\cite{KVLL} and their splitting into 
the so called instanton-dyons (anti-dyons), also known as instanton-monopoles or instanton-quarks.
 Diakonov and Petrov \cite{DP}  suggested that the back reaction of the dyons on the holonomy potential 
at low temperature may be at the origin of the disorder-order transition of the Polyakov line.
A very simple model of a de-confinement transition has been proposed by Shuryak and Sulejmanpasic~\cite{SHURYAKSUL}
through the use of dyon-antidyon ``repulsive cores".

The 
dyon-anti-dyon liquid model proposed by Diakonov and Petrov~\cite{DP} was based on (parts of) the one-loop
determinant providing the metric of the moduli spaces in BPS-protected sectors, purely selfdual or antiselfdual.
The dyon-antidyon interaction is not BPS protected and appears at the leading -- classical -- level,
related with the so called streamline configurations, the solutions of the ``gradient flow" equation.
These solutions have been  recently derived by Larsen and Shuryak \cite{LARSEN}. Their inclusion in our work \cite{LIU1}
reveals a very strong coupling of the dyons to the anti-dyons, which can however 
 be effectively reduced by screening, provided the dyon ensemble is dense enough.

Before turning to the main subject of this work which is focused on the effects of 
light quarks on the gauge topology and chiral symmetry, we will briefly mention some
important studies for the development of our work. The original discovery of 
the KvBLL instantons~\cite{KVLL} with non-trivial holonomies  is the key starting
point for assessing the role of center symmetry on the gauge topological structures.
The second important development is the assessment of the quantum weight around
the KvBLL instantons in terms of the coordinates of the instanton-dyons developed
by Diakonov and collaborators~\cite{DP,DPX}. 
%
%
%
The  dissociation of instantons into fractional constituents
 is similar to the  Berezinsky-Kosterlitz-Thouless (BKT) transition in 2-dimensional CPN models~\cite{FATEEV},
as has been advocated by Zhitnitsky  and collaborators~\cite{ZHITNITSKY} , although  substantially different in the details.

A center-symmetric (confining) phase can be compatible with an exponentially dilute regime
that is controlled semi-classically, as shown  by Unsal and Yaffe~\cite{UNSAL1}  using a double-trace
deformation of Yang-Mills action at large $N$  on $S^1\times R^3$.
A similar trace deformation was used originally in the context of two-dimensional (confining) QED with unequal
charges on $S^1\times R$~\cite{HOLGER} to analyze center symmetry and its spontaneous breaking. 
This construction was extended to QCD with adjoint fermions by Unsal~\cite{UNSALALL},
and  by Unsal and others~\cite{UNSAL} to a class of  deformed supersymmetric theories with soft supersymmetry breaking. 
 While the setting includes a compactification on a small circle, with  weak coupling and
 an exponentially  $small$ density of dyons, the minimum at the confining holonomy
  value is induced by the repulsive interaction in the dyon-anti-dyon pairs (called  
 $bions$ by the authors). 
A key role of  supersymmetry is the cancellation of the perturbative Gross-Pisarski-Yaffe-Weiss (GPYW)  holonomy potential \cite{WEISS}.
While this allows to study deconfinement transition in the very dilute regime, the major subject
to be studied in this work -- spontaneous chiral symmetry breaking -- would still be absent,
as its development would require an ensemble which is sufficiently dense.

  Let us now turn to the effects of light fermions. Key to these effects are topological index theorems, which 
  relate the topological charge of the solitons to the number of its fermionic zero modes.   
  When the ensemble of topological solitons is dense enough, the fermionic zero modes can
  collectivize and produce the so called Zero Mode Zone (ZMZ) which breaks spontaneously
 chiral symmetry. For ensemble of instantons this phenomenon has been studied in
  great detail in 1980's and 1990's, for a review see \cite{ALL}.
   Thanks to   topology, the fermionic zero modes are remarkably stable agains any smooth deformations of these
  objects, resisting tremendous amount of perturbative noise. As has been derived in the 
  ``instanton liquid model" context and many times observed 
in lattice numerical simulations,  the ZMZ states with Dirac eigenvalues in the range $| \lambda | \leq 20\, {\rm MeV}$ ,
  are crucial for the generation of the hadronic masses and properties, while being only a tiny subset of  all fermionic states (typically of the order of $10^{-4}$ in current lattice simulations).

  Since the instanton-dyons carry topological charge, they should have zero modes as well.   On the other hand, 
  for an arbitrary number of colors $N_c$ those topological charges are fractional $1/N_c$, while the
 number of zero modes must be integers. Therefore
  only some instanton-dyons may have zero modes. For general $N_c$
 and general periodicity angle of the fermions the answer is known but a bit involved. For 
 $SU(2)$ colors and physically anti-periodic fermions the twisted $L$ dyons have zero modes,
 while the usual $M$-dyons do not.     
  
Recent investigations by Shuryak and others~\cite{SHURYAK,Faccioli:2013ja}  have shown that
light fermions disorder in ensembles composed of interacting dyons and anti-dyons. Their numerical
analyses support the accumulation of zero virtuality quark states leading to the spontaneous breaking
of chiral symmetry. The effects of light quarks on deformed supersymmetric models where recently addressed
in~\cite{TIN}.

In this work we would like to follow up on our study in~\cite{LIU1} by allowing for light quarks
in the dense center symmetric phase of the dyon-anti-dyon Coulomb plasma.  
The word ``dense" is  key here, as it justifies the use of a mean-field analysis in the
characterizing the spontaneous breaking of chiral symmetry and the formation of a chiral
condensate. As our interest is now in the light quark dynamics, we will
only enforce the strong Coulomb corrections at the constraint level. One of the chief achievement of this work is
to show how the induced chiral effective Lagrangian knows about confinement. In particular, we detail the
interplay between the spontaneous breaking of chiral symmetry and center symmetry.

In section 2 we detail the color SU(2) version of the model for
$N_f=1$. By using a series of fermionization and bosonization techniques we show how the
3-dimensional effective action for the liquid can be constructed to accommodate for the light quarks.
In section 3, we show that the ground state solution supports both center symmetry and chiral condensation.
In section 4 we detail the flavor spectrum in terms of the sigma meson, the eta$^\prime$ meson which
is shown to be anomalous. 
In section 5, we explore the effects of molecular pairing of dyons and anti-dyons induced by the light quarks
near the
transition temperature and their effect on the  formation of the chiral condensate and center symmetry.
In section 6 we briefly extend the model to include many colors and flavors
and show that in the dyon-anti-dyon liquid with light quarks, the restoration of chiral symmetry occurs
simultaneously with the loss of center symmetry for $x=N_f/N_c\geq 2$.
 An estimate of the transition temperature from the center symmetric to non-symmetric
phase  is made. Our conclusions are in section 7.

\section{ Effective action with fermions}

\subsection{General setting} 

Since this is the second paper of the series, we will keep all notations consistent with the first paper~\cite{LIU1}
which should be consulted for further details. To keep the paper self-contained and before introducing the fermions,
we will summarize some essential points for the current setting.

In the semi-classical approximation, the Yang-Mills partition function is assumed to be dominated by an interacting ensemble of
instanton-dyons (anti-dyons). 
For inter-particle distances large compared to their sizes -- or a very dilute ensemble --
 both the classical interactions and the one-loop effects are Coulomb-like. At 
 distances of the order of  the particle  sizes the one-loop effects are
 encoded in the geometry of the moduli space of the ensemble. For multi-dyons  a plausible moduli space was argued starting
from the KvBLL caloron~\cite{KVLL} that has a number of pertinent symmetries, among which permutation symmetry, overall charge neutrality,
and clustering to KvBLL. Since the underlying calorons are self-dual, the induced metric on the moduli space was shown to be
hyper-Kahler.

Specifically and for a fixed holonomy $A_4(\infty)/2\omega_0=\nu \tau^3/2$ with $\omega_0=\pi T$ and $\tau^3/2$
being the only diagonal color algebra generator, the
SU(2) KvBLL instanton (anti-instanton) is composed of a pair of dyons labeled by L, M (anti-dyons by $\overline {\rm L},\overline {\rm M}$)
in the notations of~\cite{DP}. 
Generically there are $N_c-1$ M-dyons and only one twisted  L-dyon type. For the SU(2) gauge group used for most of our
discussion,  M  carries (electric-magnetic) charges $(+,+)$ and L  carries $(-,-)$, with
fractional topological charges $v_m=\nu$ and $v_l=1-\nu$,  respectively. Their corresponding actions are $S_L=2\pi v_m/\alpha_s$ and  $S_M=2\pi v_l/\alpha_s$. The M-dyons are also referred to as BPST dyons, while the L-dyons as Kaluza-Klein dyons.


With the above in mind 
the SU(2) grand-partition function  is
written as 

\bea
{\cal Z}_{1}[T]&&\equiv \sum_{[K]}\prod_{i_L=1}^{K_L} \prod_{i_M=1}^{K_M} \prod_{i_{\bar L}=1}^{K_{\bar L}} \prod_{i_{\bar M}=1}^{K_{\bar M}}\nonumber\\
&&\times \int\,\frac{f_Ld^3x_{Li_L}}{K_L!}\frac{f_Md^3x_{Mi_M}}{K_M!}
\frac{f_Ld^3y_{{\bar L}i_{\bar L}}}{K_{\bar L}!}\frac{f_Md^3y_{{\bar M}i_{\bar M}}}{K_{\bar M}!}\nonumber\\
&&\times {\rm det}(G[x])\,{\rm det}(G[y])\,\left|{\rm det}\,\tilde{\bf T}(x,y)\right|\,\,e^{-V_{D\overline D}(x-y)}\nonumber\\
\label{SU2}
\eea
Here $x_{mi}$ and $y_{nj}$ are the 3-dimensional coordinate of the i-dyon of  m-kind
and j-anti-dyon of n-kind. Here
$G[x]$ a $(K_L+K_M)^2$ matrix and $G[y]$ a $(K_{\bar L}+K_{\bar M})^2$ matrix whose explicit form are given in~\cite{DP,DPX}.
$V_{D\bar D}$ is the streamline interaction between ${\rm D=L,M}$ dyons and ${\rm \bar D=\bar L, \bar M}$ antidyons as numerically discussed
in~\cite{LARSEN}. For the SU(2) case its Coulomb asymptotic is~\cite{LIU1}

\bea
&&V_{D\overline{D}}(x-y)\rightarrow-\frac {C_D}{\alpha_s\, T}\nonumber\\
&&\times\left(\frac 1{|x_M-y_{\overline{M}}|}+\frac 1{|x_L-y_{\overline{L}}|}-\frac 1{|x_M-y_{\overline{L}}|}-\frac 1{|x_L-y_{\overline{M}}|}
\right)\nonumber\\
\label{DDXX}
\eea
The strength of the classical Coulomb interaction in (\ref{DDXX}) is $C_D/\alpha_s=2.46/\alpha_s$. At intermediate distances
$V_{D\bar D}$ is characterized by a core $a_{D\bar D}\approx 1/T$.
The key new element in the partition function (\ref{SU2}) in comparison to our previous work \cite{LIU1},  is the introduction
of the fermionic determinant ${\rm det}\,\tilde{\bf T}(x,y)$ that we will discuss further below.

The fugacities $f_{i}$ are related to the overall dyon density.
The dyon density $n_D$ could be extracted from lattice measurements of the caloron plus anti-caloron
densities at finite temperateure in unquenched lattice simulations. Following~\cite{LIU1} we define
\be
\frac{n_D}{T^3}=C\,\frac{e^{-\frac{\pi}{\alpha_s}}}{\alpha_s^2}
\label{AA0}
\ee
with $C$ a constant whose value depends on
the regularization scheme of the divergent determinant, and ultimately on the specific definition of 
$\Lambda_{\rm QCD}$. For definiteness, we will use
\be
\frac{\pi}{\alpha_s}=\frac{10}{3}\,{\rm ln}\left(\frac T{0.36\,T_c}\right)
\label{AA1}
\ee
where $10/3=11N_c/6-N_f/3$ for $N_c=2$ and $N_f=1$.
The constant inside the logarithm  has been fitted to lattice measurements of  the instanton density  for  $N_c=2$ and $N_f=0$.
In principle, it should be modified along with $T_c$, as the theory changes, e.g. $N_f=0$ to $N_f=1$.
Since we do not have such lattice data, we will only modify  the  beta function coefficient in front.

We conclude this section by addressing some limitations of the model described by (\ref{SU2}). One 
limitation is that the dyonic plasma should be dense enough  to produce sufficiently large screening masses,
as discussed in detail in \cite{LIU1}. In practice, this
limits its application to the confined phase with $T<T_c$.  The model starts to get
inapplicable at high density when  the dyons are close to the maximal packing density. Another 
limitation is that at small enough $T$ the action per dyon $8\pi^2/g^2(T) N_c$ becomes small
lthereby invalidating  the use of the semiclassical approximation. Our estimates in~\cite{LIU1}
show that the model can still be used with reasonable accuracy in the range $0.5\,T_c<T<T_c$.   

\subsection{Quark zero modes}

For quarks in the fundamental color representation, the squared Dirac equation in an external chromo-magnetic ${\bf B}$ 
and chromo-electric ${\bf E}$ field,  takes the
generic form in the chiral representation~\cite{DIRAC}

\be
\left (-\nabla^2+4\,{\bf S}\cdot ({\bf B}\mp {\bf E})\right)\varphi^\pm=0
\label{DIR}
\ee
with $i\nabla=i\partial  +A$ and     ${\bf S}^a$ the SU(2) spin generators. The signs in (\ref{DIR}) are commensurate
with chirality. In the absence of spin, scalar quarks do not
admit zero modes as $-\nabla^2$ is semi-positive. With spin, zero modes may occur  when the spin contribution is negative in (\ref{DIR}) to balance the semi-positive scalar contribution. For a self-dual object ${\bf B}={\bf E}$ and only the negative chirality
quark  can produce a zero mode state through the ``magnetic moment term"

\be
\left (-\nabla^2+4\,{\sigma}\cdot {\bf B}\right)\varphi_D^-=0
\label{DIR1}
\ee
In the dyon the last term is $\sigma\cdot {\bf B}\approx \sigma\cdot \hat{r}/\rho^2$ at
the core size $\rho\approx 1/\nu\omega_0$.
The first term in (\ref{DIR1}) is the squared kinetic energy. It is  bounded by the uncertainty principle
and of order $1/\rho^2$.  These two terms can balance by an 
anti-aligned spin contribution at the core. Hence the negative chirality state is bound in the dyon, while the
positive chirality state is bound in the anti-dyon.

The explicit fermionic zero modes of the KvBLL instanton were discussed in~\cite{KRAAN}. It was noted that for large holonomies
the zero mode is localized on one of the constituent dyon. The zero modes on individual SU(2) dyons were made explicit
in~\cite{SHURYAK}.  For SU(2)
in the center disordered phase with $\nu=1/2$ and zero winding, the M-dyon only supports fermionic zero modes with periodic
boundary condition. In the string gauge for the M-dyon, the Dirac equation for the
zero mode reads asymptotically

\be
\left(\partial_4\mp i\nu\omega_0-i\sigma \cdot \partial \right)\varphi^\pm=0
\label{DIRAC}
\ee
with $\varphi^{\pm}$  the upper and lower components respectively. The solutions are hedgehogs
with $\varphi^{\pm}=e^{-i\phi x_4/\beta}\sigma\cdot \hat{r}\,f^{\pm}(r)$,

\be
\left(\partial_r+2/r+\nu\omega_0\mp \phi/\beta\right)f^{\pm}(r)=0
\label{DIRAC1}
\ee
Normalizable solutions require $\phi/\beta<\nu\omega_0$, ruling out the anti-periodic boundary condition.
On the other hand, the L-dyon zero mode follows from the M-dyon by switching $\nu$ to $(1-\nu)$ and then
performing a time-dependent gauge transformation $U=e^{-i\omega_0 x_4 \tau_3}$. Thus the L-dyon zero mode
is anti-periodic for $\nu=1/2$.

The L- and M-zero modes associated with  the higher
winding sectors labeled by $n$ on $R^3\times S^1$~\cite{UNSAL} follow from the substitution
$\nu\rightarrow \nu+n$ in (\ref{DIRAC1}). For $\phi=(2m+1)\pi$, there are normalizable
M-zero modes if $-\frac{2n+3}{4}<m<\frac{2n-1}{4}$. For $\phi=2m^\prime\pi$, there are normalizable M-zero modes
if $-\frac{2n+1}{4}<m^\prime<\frac{2n+1}{4}$. The corresponding L-zero modes follow by gauge transformation.
Thus anti-periodic L-zero modes in a sector $n$ correspond to periodic M-zero modes in the same sector and
vice versa. Throughout, only the $n=0$ sector will be discussed for clarity.

\subsection{Fermionic determinant}

The main issue discussed in  this paper is the behavior (pairing or collectivization) of the fermionic zero modes
into what is called in the literature the ``Zero Mode Zone" (ZMZ). The
approximations used in its description follows closely the
 construction, developed for instantons and described in detail in refs~\cite{ALL}.
The fermionic determinant 
can be viewed as a sum of closed fermionic loops connecting all dyons and antidyons. Each link 
-- or hopping --
 between L-dyons and ${\rm \bar{L}}$-anti-dyons is described by the elements of the ``hopping
chiral matrix" $\tilde{\bf T}$

\begin{eqnarray}
\tilde {\bf T}(x,y)\equiv \left(\begin{array}{cc}
0&{\bf T}_{ij}\\
-{\bf T}_{ji}&0
\end{array}\right)
\label{T12}
\end{eqnarray}
with dimensionality $(K_L+K_{\bar L})^2$.
Each of the entries in ${\bf T}_{ij}$ is a  ``hopping amplitude" for a fermion between
an L-dyon and an $\bar{\rm L}$-anti-dyon,  defined via the zero mode $\varphi_D$ of the dyon and the zero mode
$\varphi_{\bar D}$ (of opposite chirality) of the anti-dyon 

\be
{\bf T}_{ij}\equiv {\bf T}(x_i-y_j)=\int  d^4z\, \varphi_{\bar D}^\dagger(z-x_i)i(\gamma \cdot \partial) \varphi_D (z-y_j)
\label{TIJ}
\ee
 The exact zero modes in the hedgehog gauge relate to the zero modes in the string gauge through
$\eta_{A\alpha}=-\varphi_\beta^A\epsilon_{\beta\alpha}$ with indices $A$ for color and $\alpha$ for spinors. They are explicitly given by
~\cite{SHURYAK}


\be
\eta_{A\alpha}=\frac{\omega_0^{\frac 32}}{2\sqrt{8\pi}} 
\frac{{\rm th} \frac{x}{2}}{\sqrt{x\,{\rm sh} x}}
\left(1-\sigma \cdot \hat{r}\right)_{A\alpha}
\label{DZERO}
\ee
with $x=\omega_0 r$. Since ${\rm Tr}(\varphi_1^\dagger\varphi_2)={\rm Tr}(\eta_1^\dagger\eta_2)$, we may substitute
(\ref{DZERO}) into (\ref{TIJ}). The Fourier transform of the result is

\be
{\bf T}(p)=\frac{\omega_0}2\left(|A_1(p)|^2+|A_0^\prime(p)\right|^2
\label{TP}
\ee
with

\be
A_n(p)=\frac{\sqrt{2\pi}}{\,\omega_0^{n+\frac 12}}\int_0^\infty
dx\,x^{n+\frac 12}\,\frac{{\rm sin}(\tilde p x)}{\tilde p x}\frac {{\rm th}\frac x2}{\sqrt{\,{\rm sh} x}}
\label{Anp}
\ee
with $\tilde p=p/\omega_0$. The two integrals  in (\ref{Anp}) for $n=0,1$ will be carried numerically for
our results below. 



Since we will be interested in the center symmetric phase of the dyon-anti-dyon ensemble we have left out 
the repulsive linear interaction between unlike dyons (anti-dyons) in the KvBLL instanton as  each sector
already assume them dissociated. This pair interaction acts as a linearly confining force in the center disordered
phase. A schematic and local pair-interaction between dyons and anti-dyons induced by the fermions will be
added to (\ref{T12}) below. 

Depending on the dyon density and locations, the determinant can either be dominated by small
(binary) loops, or very long loops connecting macroscopically large number of dyons.
The first phase  is called 
 ``molecular" and is dominated by dyon-anti-dyon clusters,   reminiscent of the molecules in the instanton ensemble~\cite{ILGEN}.
The second phase of very long loops is called ``collectivization" and leads to a nonzero quark condensate.

\subsection{Bosonic fields}

Following~\cite{DP,LIU1} the moduli
determinants in (\ref{SU2}) can be fermionized using 4 pairs of ghost fields $\chi^\dagger_{L,M},\chi_{L,M}$ for the dyons
and 4 pairs of ghost fields $\chi^\dagger_{{\bar L},{\bar M}},\chi_{{\bar L},{\bar M}}$ for the anti-dyons. The ensuing Coulomb factors from the determinants are then bosonized using 4 boson fields $v_{L,M},w_{L,M}$ for the dyons and similarly for
the anti-dyons.  The result is a doubling of the 3-dimensional free actions obtained in~\cite{DP}

\bea
&&S_{1F}[\chi,v,w]=-\frac {T}{4\pi}\int d^3x\nonumber\\
&&\left(|\nabla\chi_L|^2+|\nabla\chi_M|^2+\nabla v_L\cdot \nabla w_L+\nabla v_M\cdot \nabla w_M\right)+\nonumber\\
&&\left(|\nabla\chi_{\bar L}|^2+|\nabla\chi_{\bar M}|^2+\nabla v_{\bar L}\cdot \nabla w_{\bar L}+\nabla v_{\bar M}\cdot \nabla w_{\bar M}\right)
\label{FREE1}
\eea
For the interaction part $V_{D\bar D}$, we note that
the pair Coulomb interaction in (\ref{SU2}) between the dyons and anti-dyons can also be bosonized using
standard tricks~\cite{POLYAKOV,KACIR}  in terms of $\sigma$ and $b$ fields. We note that $\sigma$ and $b$
are the un-Higgsed long range U(1) parts of the original magnetic field $F_{ij}$ and electric potential
$A_4$ (modulo the holonomy) respectively.  As a result each dyon species acquire additional
fugacity factors such that

\be
M:e^{-b-i\sigma}\,\,\,\,\, L:e^{b+i\sigma}\,\,\,\,\, \bar M: e^{-b+i\sigma}\,\,\,\,\, \bar L :e^{b-i\sigma}
 \ee
Note that these assignments are consistent with those suggested in~\cite{UNSAL,SHURYAKSUL} using different arguments.
As a result there is an additional contribution to the free part (\ref{FREE1})

\be
S_{2F}[\sigma, b]=\frac T{8} \int d^3x\, \left(\nabla b\cdot\nabla b+ \nabla\sigma\cdot\nabla\sigma\right)
\label{FREE2}
\ee
and the interaction part is now

\bea
&&S_I[v,w,b,\sigma,\chi]=-\int d^3x \nonumber\\
&&e^{-b+i\sigma}f_M\left(4\pi v_m+|\chi_M    -\chi_L|^2+v_M-v_L\right)e^{w_M-w_L}+\nonumber\\
&&e^{+b-i\sigma}f_L\left(4\pi v_l+|\chi_L    -\chi_M|^2+v_L-v_M\right)e^{w_L-w_M}+\nonumber\\
&&e^{-b-i\sigma}f_{\bar M}\left(4\pi v_{\bar m}+|\chi_{\bar M}    -\chi_{\bar L}|^2+v_{\bar M}-v_{\bar L}\right)e^{w_{\bar M}-w_{\bar L}}+\nonumber\\
&&e^{+b+i\sigma}f_{\bar L}\left(4\pi v_{\bar l}+|\chi_{\bar L}    -\chi_{\bar M}|^2+v_{\bar L}-v_{\bar M}\right)e^{w_{\bar L}-w_{\bar M}}
\label{FREE3}
\eea
without the fermions. We now show the minimal modifications to (\ref{FREE3}) when the fermionic determinantal
interaction is present.

\subsection{Fermionic fields}

The determinant for the hopping fermionic zero mode can be fermionized using standard methods. For that,
each entry ${\bf T}(x-y)$ in (\ref{SU2})  can be viewed as a cross two-body
dyon-anti-dyon hopping matrix with a two-body inverse ${\bf T}{\bf G}={\bf 1}$. To fermionize the determinant
we define the additional Grassmanians $\chi=(\chi^i_1,\chi^j_2)^T$ with $i,j=1,.., K_{L,\bar L}$ and

\be
\left|{\rm det}\,\tilde{\bf T}\right| =\int   D[\chi]\,\, e^{\,\chi^\dagger \tilde {\bf T} \, \chi}
\label{TDET}
\ee
We can re-arrange the exponent in (\ref{TDET}) by defining  a Grassmanian source $J(x)=(J_1(x),J_2(x))^T$ with

\be
J_1(x)=\sum^{K_L}_{i=1}\chi^i_1\delta^3(x-x_{Li})\nonumber\\
J_2(x)=\sum^{K_{\bar L}}_{j=1}\chi^j_2\delta^3(x-y_{\bar L j})
\label{JJ}
\ee
and by introducing 2 additional fermionic fields  $ \psi(x)=(\psi_1(x),\psi_2(x))^T$. Thus

\be
e^{\,\chi^\dagger \tilde {\bf T}\,\chi}=\frac{\int D[\psi]\,{\rm exp}\,(-\int\psi^\dagger \tilde {\bf G}\, \psi +
\int J^\dagger \psi + \int\psi^\dagger J)}{\int d
D[\psi]\, {\rm exp}\,(-\int \psi^\dagger \tilde {\bf G} \,\psi) }
\label{REFERMIONIZE}
\ee
with $\tilde{\bf G}$ a $2\times 2$ chiral block matrix

\begin{eqnarray}
 \tilde {\bf G}= \left(\begin{array}{cc}
0&{\bf G}(x,y)\\
-{\bf G}(x,y)&0
\end{array}\right)
\label{GG}
\end{eqnarray}
with entries ${\bf TG}={\bf 1}$. The Grassmanian source contributions in (\ref{REFERMIONIZE}) generates a string
of independent exponents for the L-dyons and $\bar{\rm L}$-anti-dyons

\begin{eqnarray}
\prod^{K_L}_{i=1}e^{\chi_1^i\dagger \psi_{1}(x_{Li})+\psi_1^\dagger(x_{Li})\chi_1^i}\nonumber \\ \times
\prod^{K_{\bar L}}_{j=1}e^{\chi_2^j\dagger \psi_{2}(y_{\bar L j})+\psi_2^\dagger(y_{\bar L j})\chi_2^j}
\label{FACTOR}
\end{eqnarray}
The Grassmanian integration over the $\chi_i$ in each factor in (\ref{FACTOR}) is now readily done to yield

\be
\prod_{i}[-\psi_1^\dagger\psi_1(x_{Li})]\prod_j[-\psi_2^\dagger\psi_2(y_{\bar L j})]
\label{PLPR}
\ee
for the L-dyons and $\bar {\rm L}$-anti-dyons.
The net effect of the additional fermionic determinant in (\ref{SU2}) is to shift the L-dyon
and $\bar{\rm L}$-anti-dyon fugacities in (\ref{FREE3}) through

\bea
f_L\rightarrow -f_L\psi_1^\dagger\psi_1\equiv -f_L\psi^\dagger\gamma_+\psi\nonumber\\
f_{\bar L}\rightarrow -f_{\bar L}\psi_2^\dagger\psi_2\equiv -f_{\bar L}\psi^\dagger\gamma_-\psi
\label{SUB}
\eea
where we have now identified the chiralities through $\gamma_\pm=(1\pm \gamma_5)/2$.
The fugacities $f_{M,\bar M}$ are left unchanged since they do not develop zero modes.

\subsection{Resolving the constraints}

In terms of (\ref{FREE1}-\ref{FREE3})  and the substitution
(\ref{SUB}), the dyon-anti-dyon partition function (\ref{SU2})
for $N_f=1$ can be exactly re-written as an interacting
effective field theory in 3-dimensions,

\bea
{\cal Z}_{1}[T]\equiv &&\int D[\psi]\,D[\chi]\,D[v]\,D[w]\,D[\sigma]\,D[b]\,\nonumber\\&&\times
e^{-S_{1F}-S_{2F}-S_{I}-S_\psi}
\label{ZDDEFF}
\eea
with the additional $N_f=1$ chiral fermionic contribution $S_\psi=\psi^\dagger\tilde{\bf G}\,\psi$.
In the presence of the fermionic fields $\psi$ and the
screening fields $\sigma, b$   the 3-dimensional effective field theory (\ref{ZDDEFF})
is not  integrable. Simple approximation schemes will be developed to address this
effective action.

Note that the effective action in (\ref{ZDDEFF}) is linear in the $v_{M,L,\bar M,\bar L}$.
These are auxiliary fields that integrate  into delta-function constraints.
However and for convenience, it is best to shift away
the $b,\sigma$ fields from (\ref{FREE3}) through

\be
&&w_M-b+i\sigma\rightarrow w_M\nonumber\\
&&w_{\bar M}-b-i\sigma\rightarrow w_{\bar M}
\label{SHIFT}
\ee
which carries unit Jacobian and no anomalies, and recover them in the pertinent arguments of the delta function constraints as

\bea
&&-\frac{T}{4\pi}\nabla^2w_M+f_M  e^{w_M-w_L}\nonumber\\&&
-f_L\psi^\dagger\gamma_+\psi\,\e^{w_L-w_M}=\frac {T}{4\pi}\nabla^2(b-i\sigma)\nonumber\\
&&-\frac{T}{4\pi}\nabla^2w_L-f_Me^{w_M-w_L}
\nonumber\\&&
+f_L\psi^\dagger\gamma_+\psi\,\e^{w_L-w_M}=0
\label{DELTA}
\eea
and similarly for the anti-dyons.
To proceed further the formal classical solutions to the constraint equations or $w_{M,L}[\sigma, b]$
should be inserted back into the 3-dimensional effective action. As in~\cite{DP} we observe
 that the classical solutions
to (\ref{DELTA}) can be used to integrate the $w^\prime$s in (\ref{ZDDEFF}) to one loop. The resulting bosonic determinant
cancels against the fermionic determinant after also integrating over the $\chi^\prime$s in (\ref{ZDDEFF}). The result is

\bea
{\cal Z}_{1}[T]=\int D[\psi]\,D[\sigma]\,D[b]\,e^{-S}
\label{ZDDEFF1}
\eea
with the 3-dimensional effective action

\bea
S=&&S_F[\sigma, b]+\int d^3x\,\psi^\dagger \tilde{\bf G} \psi\\
&&-4\pi f_M v_m\int d^3x\,( e^{w_M-w_L}+e^{w_{\bar M}-w_{\bar L}} ) \nonumber \\
&& +4\pi f_Lv_l\int d^3x\,\psi^\dagger\gamma_+\psi\,e^{w_L-w_M}\nonumber\\
&&+4\pi f_{\bar L}v_l\int d^3x \,\psi^\dagger \gamma_-\psi\,e^{w_{\bar L}-w_{\bar M}}\nonumber
\label{NEWS}
\eea
Here $S_F$ is $S_{2F}$ in (\ref{FREE2}) plus additional contributions resulting from the $w_{M,L}(\sigma, b)$ solutions
to the constraint equations (\ref{DELTA}) after their insertion back.  This procedure for the linearized approximation of the constraint
was discussed in~\cite{LIU1} for the case without fermions.

\section{SU(2) QCD with one quark flavor}

To analyze the ground state and the fermionic fluctuations we  bosonize the fermions
in (\ref{ZDDEFF1})  by introducing two delta-functions and re-exponentiating them

\bea
{\cal Z}_{1}[T]=\int &&D[\psi]\,D[\sigma]\,D[b]\,D[\Sigma]\,D[\Sigma_5] \,D[\lambda]\,D[\lambda_5]\,\nonumber\\
&&\times \,\,\,e^{-S+\int d^3x\,i\lambda(\psi^\dagger\psi+\Sigma)+\int d^3x\,i\lambda_5(\psi^\dagger i\gamma_5\psi+\Sigma_5)}\nonumber\\
\label{ZDDEFF2}
\eea
The ground state is parity even so that $f_{L,M}=f_{\bar L, \bar M}$ and $\Sigma_5=0$.
By translational invariance, the SU(2) ground state corresponds to constant $\sigma, b, \Sigma$.
The classical solutions to the constraint equations (\ref{DELTA}) are also constant

\be
(e^{w_M-w_L})_0=\sqrt{f_L\Sigma/2f_M }
\label{SU2SOL}
\ee
and similarly for the anti-dyons.

\subsection{Effective potential}

The effective potential ${\cal V}$ for constant fields follows from (\ref{ZDDEFF2})
by enforcing the delta-function constraint (\ref{NEWS}) and parity

\bea
-{\cal V}/\mathbb{V}_3=&&+i\lambda\Sigma\nonumber\\&&+4\pi f_M v_m\,( e^{w_M-w_L}+e^{w_{\bar M}-w_{\bar L}} ) \nonumber \\
&& +2\pi f_Lv_l\,\Sigma\,(e^{w_L-w_M}+e^{w_{\bar L}-w_{\bar M}})
\label{POT}
\eea
with $\mathbb{V}_3$ the 3-volume.
For fixed holonomies $v_{m,l}$, the constant $w^\prime$s are real by (\ref{DELTA})
as all right hand sides vanish, and the extrema of (\ref{POT}) occur for

\bea
e^{w_M-w_L}=\pm \sqrt{\Sigma f_Lv_l/2f_Mv_m}\nonumber\\
e^{w_{\bar M}-w_{\bar L}}=\pm \sqrt{\Sigma f_{L}v_{\bar l}/2f_Mv_{\bar m}}
\label{EXT}
\eea
(\ref{EXT}) are  consistent with  (\ref{SU2SOL}) only if
$v_l=v_m=1/2$ and $v_{\bar l}=v_{\bar m}=-1/2$. That is for confining holonomies or a center
symmetric ground state. Thus

\be
- {\cal V}/\mathbb{V}_3=i\lambda\Sigma +8\pi \sqrt{f_L f_M\Sigma/2}
\label{SU2POT}
\ee
We note that for $\Sigma=0$ there are no solutions to the extrema equations. The holonomies are
no longer constrained to the center symmetric state. Since $\Sigma=0$ means a zero chiral condensate
(see below), we conclude that in this model of the  dyon-anti-dyon liquid with light quarks, chiral symmetry restoration and
the loss of center symmetry occur simultaneously.



For the vacuum solution, the auxiliary field $\lambda$ is also a constant.
The fermionic fields in (\ref{ZDDEFF2}) can be integrated out. The result is
a new contribution to the potential (\ref{SU2POT})

\bea
- {\cal V}/\mathbb{V}_3\rightarrow&& +i\lambda\Sigma +8\pi \sqrt{f_L f_M\Sigma/2}
\nonumber\\&&+\int \frac{d^3p}{(2\pi)^3}{\rm ln}\,\left(1-\lambda^2{\bf T}^2(p)\right)
\label{SU2POT1}
\eea
The saddle point  of (\ref{SU2POT1}) in $\Sigma$ is solution to

\be
i\lambda+\frac{\alpha}{\sqrt{\Sigma}}=0\rightarrow \lambda=\frac{\alpha}{\sqrt{\Sigma}}
\label{L1}
\ee
after the substitution $\lambda \rightarrow i\lambda$ with $\alpha=4\pi \sqrt{f_Lf_M/2}$. 
Inserting (\ref{L1}) into the effective potential (\ref{SU2POT1}) yields

\be
- {\cal V}/\mathbb{V}_3=\frac{\alpha^2}{\lambda}+\int \frac{d^3p}{(2\pi)^3}{\rm ln}\,\left(1+\lambda^2{\bf T}^2(p)\right)
\label{E1}
\ee
The saddle point for $\lambda$ is 
\be
\frac{\alpha^2}{2\lambda}=\int \frac{d^3p}{(2\pi)^3} \frac{\lambda^2{\bf T}^2(p)}{1+\lambda^2{\bf T}^2(p)} \equiv V_0
\label{S1}
\ee
It is readily checked that (\ref{S1}) enforces the true minimum condition $d({\cal V}/\mathbb{V}_3)=0$. 
From (\ref{Anp}) we note that $\lambda{\bf T}(p)\approx \lambda\omega^4_0/p^{6}$ falls rapidly 
with momentum for $p>p_{\rm max}$ with $p_{\rm max}^3\equiv \omega^2_0\sqrt{\lambda}$. A simple solution to (\ref{S1}) follows from the condition 
$\lambda{\bf T}(0)\gg 1$, i.e. 

\be
V_0\approx p_{\rm max}^3\equiv \omega^2_0\sqrt{\lambda}
\label{V0}
\ee
The precise value of $V_0$ is not important as it will be traded for the dyon density below.
Note that for the opposite case of $\lambda{\bf T}(0)\ll 1$ we have $V_0\approx \lambda^2/\omega_0$.
This is the dilute dyonic density limit which is not our case. The dyon ensemble 
in the center symmetric phase is dense~\cite{LIU1}. In terms of (\ref{V0}) all equations can be solved
analytically. However, we have checked that their accuracy is limited. All the analysis to follow  will be carried exactly
without these estimates.

\subsection{Gap equation}

\begin{figure}[h!]
  \begin{center}
  \includegraphics[width=7cm]{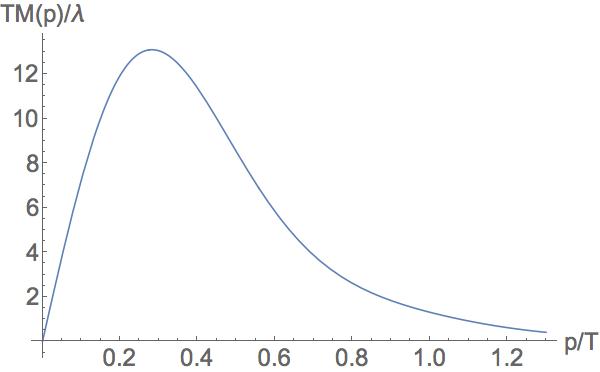}
   \caption{The momentum dependent quark constituent mass $TM(p)/\lambda$ versus $p/T$. }
     \label{fig_mass}
  \end{center}
\end{figure}

The free energy depends on two parameters, the mean values of $\lambda$ and $\Sigma$ fields,
which should be chosen at the minimum of it. The equations following from vanishing first derivatives
 are known in literature as the ``gap equations".
It is useful, to recast it in terms of the integral $V_0$ defined above. In particular, we have

\be
\Sigma=\frac{4V_0^2}{\alpha^2} =\frac{2V_0}{\lambda}
\label{SIGMA0}
\ee
while the effective potential (\ref{E1}) is 

\be
- {\cal V}/\mathbb{V}_3=2V_0+\int \frac{d^3p}{(2\pi)^3}\,{\rm ln}\left(1+\frac{M^2(p)}{p^2}\right)
\ee
We have introduced the momentum-dependent constituent quark mass $M(p)$ as

\be
M(p)=\lambda\,p {\bf T}(p)=\frac{\alpha^2}{2V_0}p{\bf T}(p)
\label{MASS}
\ee
which is seen to vanish linearly at $p/\omega_0\ll 1$ and as $1/p^2$ for $p/\omega_0\gg 1$.
In Fig.~\ref{fig_mass} we show the behavior of dimensionless mass ratio $TM(p)/\lambda$
as a function of $p/T$.  (\ref{MASS}) through (\ref{S1}) obeys the gap equation
for the $\lambda$ parameter
\be
\int \frac{d^3p}{(2\pi)^3}\frac{M^2(p)}{p^2+M^2(p)}= \frac{n_D}4
\label{GAPX}
\ee
related the integral we called $V_0$ to the dyonic density $n_D$.
So given $n_D$, the solution to the gap equation (\ref{DM}) 
fixes $\lambda$  and thus the quark constituent mass $M(p)$ and  therefore, through the delta-function constraint
in (\ref{ZDDEFF2}), the value of $\Sigma$.

In our approach the $n_D$ is not an external input, but should itself be calculated from the
derivatives of the free energy, e.g. the M-dyon density is 

\be
n_M=\frac 12 \frac{\partial ({-\cal V}/{\mathbb{V}_3})}{\partial {\rm ln} f_M}=
\int \frac{d^3p}{(2\pi)^3}\frac{M^2(p)}{p^2+M^2(p)}=V_0
\label{DM}
\ee
Since $n_D=n_M+n_L+n_{\bar M}+n_{\bar L}=4n_M$ as all partial dyonic densities are
equal in the confined phase, we have $V_0=n_D/4$.  

\subsection{Chiral condensate}

The non-vanishing of  $\Sigma$ signals the non-vanishing of the chiral condensate $\left<\bar q q\right> $
and therefore the spontaneous breaking of chiral symmetry .  Standard demonstration of that is done via introduction of a nonzero but small light quark mass $m$, which
changes (\ref{PLPR})
to
\be
\prod_{i}[-\psi_1^\dagger\psi_1(x_{L i})+m]\prod_j[-\psi_2^\dagger\psi_2(y_{\bar L j})+m]
\label{PLPRX}
\ee
A rerun of the bosonization scheme with (\ref{PLPRX}) shows that only one contribution
in  (\ref{SU2POT}) is now shifted

\be
8\pi \sqrt{f_L f_M\Sigma/2}\rightarrow 8\pi \sqrt{f_L f_M(\Sigma/2+m)}
\label{SHIFT1}
\ee
changing the saddle point solutions (\ref{L1}) to

\be
\lambda=\frac{\alpha}{\sqrt{\Sigma+2m}}
\label{LM}
\ee
and (\ref{S1}) to

\be
\frac{\alpha^2}{2\lambda}-m\lambda=\int \frac{d^3p}{(2\pi)^3}\frac{\lambda^2{\bf T}^2(p)}{1+\lambda^2{\bf T}^2(p)}
\label{GLM}
\ee
The effective potential is now

\be
- {\cal V}/\mathbb{V}_3=\frac{\alpha^2}{\lambda} +2m\lambda+\int \frac{d^3p}{(2\pi)^3}{\rm ln}\left(1+\lambda^2{\bf T}^2(p)\right)
\label{VEFFM}
\ee
Inserting (\ref{VEFFM}) in the general definition of the chiral condensate in the saddle point approximation

\be
\frac{\left<\bar q q\right>}T=\frac{\partial ({\cal V}/\mathbb{V}_3)}{\partial m}
\ee
and using the gap equation we obtain 

\be
\frac{\left<\bar q q\right>}T=-2\lambda
\label{CC0}
\ee
We have used that $\alpha$ is independent of $m$ and that the contribution multiplying
$\partial \lambda/\partial m$ is zero 
thanks to the gap equation. 
In the chiral limit, $\lambda$ is fixed by the
solution of the gap-equation (\ref{GAPX}) for the constituent quark mass and the dyon density
$n_D$. It is therefore an implicit function  of $n_D$, i.e. $\lambda\equiv \lambda[n_D]$.

\begin{figure}[h!]
  \begin{center}
  \includegraphics[width=7cm]{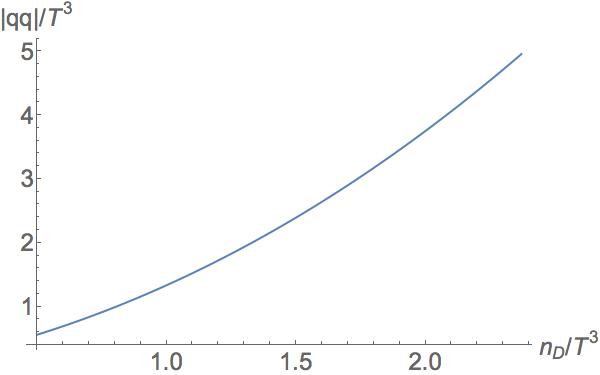}
   \caption{The absolute value of the (dimensionless) quark condensate $|\left<\bar q q\right>|/T^3$ versus the (dimensionless) dyon density $n_D/T^3$. }
     \label{fig_qqbar}
  \end{center}
\end{figure}

The use of (\ref{AA0}) into (\ref{GAPX})  leads to 
only numerical results for $\lambda$ and thus $M(p)$. 
In Fig.~\ref{fig_qqbar} we show the behavior of the absolute value of the quark
condensate $|\left<\bar q q\right>|/T^3$  versus the dyon density $n_D/T^3$. Note that
$|\left<\bar q q\right>|/T^3$ decreases with decreasing dyon density.   In this range, a best fit gives

\be
\frac{|\left<\bar q q\right>|}{T^3}\approx 1.25\left(\frac{n_D}{T^3}\right)^{1.63}
\label{QFIT}
\ee
For analytical estimates, we note that we can always
select a temperature in the range $0.5<T_0<T_c$ for which the chiral condensate is 
$|\bar q q|/T_0^3=1$ at that temperature. From (\ref{CC0}) we have $\lambda_0=T_0^2/2$
in the chiral limit. Inserting this value
for $M(p)$ in (\ref{GAPX})  shows that the corresponding dyon density at $T=T_0$ is $n_0/T^3_0=0.80$.

\subsection{Screened Polyakov lines}

As we noted earlier in (\ref{EXT}),  the spontaneous breaking of chiral symmetry with a finite value of 
$\Sigma/2$, still preserves center symmetry with $v_l=v_m=1/2$. However, strict confinement is lost because of screening.
Heavy fundamental color charges are now screened by the light constituent quarks through the formation of
tightly bound heavy-light and colorless mesons. The bound mesons are blind to the $Z_2$ center and thus
to the holonomies, with now

\bea
\left<L(x)\right>\approx &&e^{-\beta\left({\Sigma/2+m}+{\cal O}(\alpha_s)\right)}
\label{HL}
\eea
The  ${\cal O}(\alpha_s)$ contribution in (\ref{HL}) is UV sensitive and requires a specific subtraction.
Using (\ref{CC0}) together with (\ref{SIGMA0})  and (\ref{GAPX}) where $N_c=2$, we can recast  (\ref{HL}) in the
chiral limit  into the generic relation

\be
{n_D}\approx N_c \left<\bar q q\right>{\rm ln}\left({\left<L(x)\right>}\right)\left(1+{\cal O}(\alpha_s)\right)
\label{L0}
\ee
for the dyonic density in the range  $0.5<T<T_c$. (\ref{L0}) provides for an independent estimate of
the dyon density in unquenched QCD.
Finally, we  note that for large separation, the correlation of two Polyakov lines clusters  

\be
\left<L^\dagger(x)L(0)\right>\approx \left|\left<L(0)\right>\right|^2\approx e^{-\beta\left(\Sigma+2m+{\cal O}(\alpha_s)\right)}
\label{HL1}
\ee
with a vanishing of the electric string tension  due to  light quark screening.

\section{Mesonic spectrum}

The stability of the vacuum solution with $N_f=1$ can be tested by
fluctuating in the  fermionic channel which consists of both a scalar
$\sigma$ meson and a pseudo-scalar $\eta^\prime$ meson. Both are
massive. The former through the spontaneous breaking of chiral symmetry
with finite $\Sigma$, while the latter through the U$_A(1)$ anomaly
with a finite topological susceptibility.

\subsection{Sigma meson}

A simple way to probe the scalar spectrum is to note that in the spontaneously broken state,
the fermion kinetic contribution in (\ref{DELTA}) is now

\begin{eqnarray}
 \left(\begin{array}{cc}
0&{\bf G}(x,y)\\
-{\bf G}(x,y)&0
\end{array}\right)\rightarrow 
\left(\begin{array}{cc}
i\lambda{\bf 1}_{xy}&{\bf G}(x,y)\\
-{\bf G}(x,y)&i\lambda{\bf 1}_{xy}
\end{array}\right)
\label{GGMASS}
\end{eqnarray}
with ${\bf 1}_{xy}=\delta^3(x-y)$ .The scalar meson in the long wavelength limit
can be identified with the fluctuations in the chiral condensate through
$i\lambda$ in (\ref{GGMASS})  or

\be
i\lambda\rightarrow i\lambda_0+i\delta\lambda\equiv i\lambda_0\left(1+\frac{\sigma_s}{f_s}\right)
\label{LS1}
\ee
with $\lambda_0=-\left<\bar q q\right>/2T$ in the chiral limit. Fluctuations in $\lambda$ induce also fluctuations in $\Sigma$.
Thus, consistency requires

\be
\Sigma\rightarrow \Sigma_0+\delta \Sigma
\label{SCALARS}
\ee
Inserting (\ref{LS1}) into (\ref{GGMASS}) and (\ref{SCALARS}) into the delta-constraint (\ref{ZDDEFF2})
allow for a derivation of the effective action for the fluctuating parts $\delta\Sigma, \delta\lambda$. The linear
contributions are zero by the saddle point equations. So the net contributions are quadratic and higher. In
leading quadratic order $\delta\Sigma, \delta\lambda$ mix,

\bea
{\bf S}_2[\delta \lambda,\delta \Sigma]=&&-\frac 12 \int\frac{d^3p}{(2\pi)^3}(-2i)\delta\Sigma(p)\delta \lambda(-p) \\
&&-\frac{1}{2}\int\,\frac{d^3p}{(2\pi)^3} \frac{n_D}{4\Sigma_0^2}\delta\Sigma(p)\delta \Sigma(-p)\nonumber\\
&&-\frac{1}{2}\int\,\frac{d^3p}{(2\pi)^3}\,\delta\lambda(p)\,{\bf G}^{-1}_s(p)\,\delta \lambda(-p)\nonumber 
\label{SCA1}
\eea
with

\be
{\bf G}^{-1}_s(p)=\int\frac{d^3q}{(2\pi)^3}\,
\frac{2(-\lambda_0^2+{\bf G}(q^2){\bf G}(p+q)^2)}{({\bf G}^2(q^2)+\lambda_0^2)({\bf G}^2(q+p)^2)+\lambda_0^2)}\nonumber\\
\label{INT}
\ee
To undo the mixing in (\ref{SCA1})  we solve for $\delta\Sigma$ to leading order 

\be
-2i\delta \lambda -\frac{n_D}{2\Sigma_0^2}\delta \Sigma=0
\ee
and insert it back into (\ref{SCA1}) to give finally the quadratic action for the scalar meson

\be
{\bf S}_2[\sigma_s]=+\frac{1}{2f_s^2}\int\frac{d^3p}{(2\pi)^3}\sigma_s(p)
\Delta_+(p)\sigma_s(-p)\nonumber \\
\label{SCA1X}
\ee
with

\be
\label{KER}
\Delta_+(p)=\lambda_0^2{\bf G}^{-1}_s(p)+{n_D}
\ee
The kernel in (\ref{KER}) can be further reduced using the gap equation for $n_D$. Thus

\be
\label{MIX2}
\Delta_+(p)=
\frac{n_D}{2}+\int \frac{d^3q}{(2\pi)^3}\frac{(M_+q_-+M_-q_+)^2}{(M_+^2+q_+^2)(M_-^2+q_-^2)}\nonumber \\
\ee
Here, $M_{\pm }=M(q_{\pm})$ and $p_{\pm}=q\pm{p}/{2}$, where $M(q)$  is the running constituent mass in (\ref{MASS}).
For $N_f=1$ we have mixing between the scalar as a $\bar q q$ quark state and the scalar glueball. The $n_D/2$
contribution in (\ref{MIX2}) is just the mixing contribution, while the second contribution is clearly the $\bar q q$ quark
bubble contribution. A similar mixing in the scalar sector 
was observed in the instanton liquid model of the QCD vacuum~\cite{KACIR}.

\begin{figure}[h!]
  \begin{center}
  \includegraphics[width=7cm]{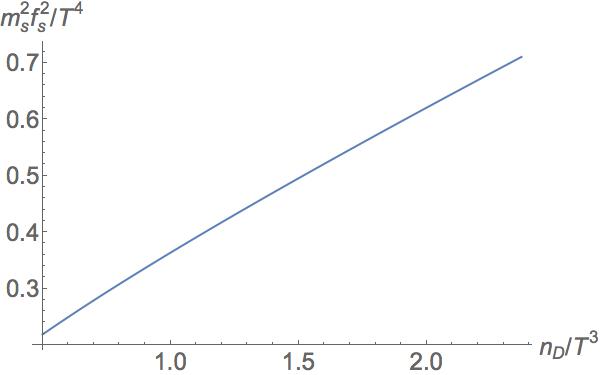}
   \caption{$m_s^2f_s^2/T^4$ versus the dyon density $n_D/T^3$. }
     \label{fig_ms2fs2}
  \end{center}
\end{figure}

A comparison of the small momentum expansion of (\ref{SCA1X}) after subtraction of the $n_D/2$ glue-mix,
yields the canonical scalar action in x-space

\be
{\bf S}_2[\sigma_s]\equiv +\frac{1}{2T}\int d^3x\,\left(|\nabla\sigma_s|^2+m_s^2{\sigma_s}^2\right)
\ee
with

\bea
m_s^2f_s^2 =T\int \frac{d^3q}{(2\pi)^3}\frac{4q^2M^2(q)}{(q^2+M^2(q))^2}
\label{SIGMA}
\eea
In Fig.~\ref{fig_ms2fs2} we show the behavior of $m_s^2f_s^2/T^4$ from (\ref{SIGMA}) as a function of the
scaled dyon density $n_D/T^3$. Increasing density amounts to lower temperature, with a transition density
expected around 1.

\subsection{eta$^\prime$ meson}

\begin{figure}[h!]
  \begin{center}
  \includegraphics[width=7cm]{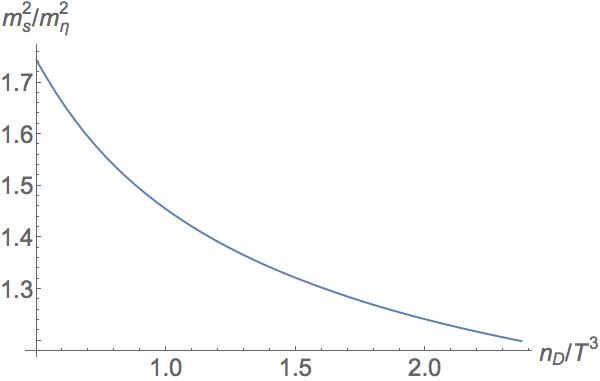}
   \caption{The mass ratio squared, for scalar to pseudo-scalar mesons  $m_s^2/m_\eta^2$, versus the (dimensionless) dyon density $n_D/T^3$, assuming $f_s\approx f_\eta$ (see text).}
     \label{fig_ms2mp2}
  \end{center}
\end{figure}

To generate the effective quadratic action for the $\eta'$ meson,  we need to fluctuate asymmetrically around the
chiral condensate in (\ref{GGMASS})

\be
i\lambda \rightarrow i\lambda_{\pm }\equiv i \lambda_0\left(1\pm \frac{i\eta}{\sqrt{2}f_\eta}\right)
\label{LAMBDAETA}
\ee
A re-run of the preceding arguments for the scalar meson yields

\be
{\bf S}_{2}[\eta]=-\frac{1}{2f_{\eta}^2}\int \eta(p)\Delta_{-}(p)\eta(-p)
\ee
with

\be
\label{KERETA}
\Delta_-(p)=\frac{\alpha^2}{\lambda_0^2}-\int\frac{d^3q}{(2\pi)^3}\frac{M_+M_-(M_+M_-+q_+q_-)}{(q_+^2+M_+^2)(q_-^2+M_-^2)}
\ee
for arbitrary current mass $m$. Using the gap equation  for $n_D$ for non-zero $m$, we may further reduce (\ref{KERETA}) into

\be
\label{KERETA1}
\Delta_-(p)=2m\lambda+\frac{n_D}{4}+\frac 12 \int \frac{d^3q}{(2\pi)^3}\frac{(q_+M_--q_-M_+)^2}{(q_+^2+M_+^2)(q_-^2+M_-^2)}\nonumber \\
\ee
Since $T\Delta_-(0)\equiv f_\eta^2m_\eta^2$, it follows that

\be
\label{WITVEZ}
f_{\eta}^2m_{\eta}^2=-{m<\bar q q>}+{\chi_T}
\ee
with $\chi_T=Tn_D/4$.
The first contribution in (\ref{QUAD2}) is the Gell-Mann-Oakes-Renner contribution to
the $\eta^\prime$ mass as a would-be Goldstone boson,
while the second contribution is a Witten-Veneziano contribution-like. It suggests an
unquenched topological susceptibility of $\chi_T=T n_D/4$ as opposed to the quenched topological 
susceptibility~\cite{LIU1}. In the chiral limit with $m=0$, 

\be
f_{\eta}^2m_{\eta}^2\approx \chi_T\equiv \frac{Tn_D}{4}
\ee
Since the topological susceptibility for dyons $\chi_T\approx {\cal O}(N_c^0)$ and $f_\eta^2\approx {\cal O}(N_c)$,
the $\eta^\prime$ mass is seen to vanish at large $N_c$. In Fig.~\ref{fig_ms2mp2} we display the ratio of the squared scalar
to pseudo scalar mass as a function of the scaled dyon density $n_D/T^3$ assuming $f_s\approx f_\eta$
by chiral symmetry. The ratio decreases with increasing dyon density 
or lower temperature. This ratio may be compared to the value in the QCD vacuum, i.e. $T=0,N_c=3,N_f=2+1$, which is
about 1/2.
Going in the opposite direction, to smaller densities, note that 
the expected phase transition density  is  around 1.

A simple but approximate understanding of (\ref{WITVEZ}) follows by noting that 
(\ref{NEWS}) is U(1)$_V$ symmetric but upsets U(1)$_A$ symmetry
through the fermionic contributions.
Under U(1)$_A$ with $\psi\rightarrow e^{i\gamma_5\theta/2}\psi$
the fermionic contributions in (\ref{POT}) change


\be
+2\pi f_Le^{i\theta}v_l\,\psi^\dagger\gamma_+\psi\,e^{w_L-w_M}
+2\pi f_{\bar L}e^{-i\theta}
v_l\psi^\dagger \gamma_-\psi\,e^{w_{\bar L}-w_{\bar M}}\nonumber\\
\ee
This amounts to shifting $f_{L,\bar L}\rightarrow f_{L,\bar L}e^{\pm i\theta}$ in the parity symmetric effective
potential (\ref{SU2POT1}). 
An estimate of mass of the $\eta^\prime$ follows by identifying $\theta/2\rightarrow \eta/f_\eta/\sqrt{2}$ with a constant
$\eta^\prime$ field. So the $\eta^\prime$ mass is related to the topological susceptibility $\chi_T$.
Specifically, the pertinent contribution from the effective potential (\ref{SU2POT1}) is now

\be
- {\cal V}/\mathbb{V}_3\rightarrow  4\pi \left({f_L f_M(\Sigma/2+m)}\right)^{1/2}\,{\rm cos}\left(\sqrt{2}\,{\eta}/{f_\eta}\right)
\label{EFF6}
\ee
where we have retained the small quark mass $m$. Expanding (\ref{EFF6})  yields
the quadratic contribution

\be
-\frac T2 \left( \frac{m\,2\alpha}{\sqrt{\Sigma} f_\eta^2}+\frac{2\alpha\sqrt{\Sigma}}{f_\eta^2}\right)\,\eta^2
\label{QUAD2}
\ee
Recall that  $\alpha=4\pi\sqrt{f_Mf_L/2}$. 
From (\ref{CC0}) we have $2\alpha/\Sigma=-\left<\bar q q\right>$ and from 
(\ref{SIGMA0}) we have $\alpha\sqrt{\Sigma}=2V_0=n_D/2$. The squared $\eta^\prime$ mass follows
as in (\ref{WITVEZ})  but with an incorrect $4\chi_T$ contribution.

\section{Dyon pairing through fermion exchanges}

Above the temperature of the chiral phase transition, $T>T_\chi$,
 there is no quark condensate, and dyons and anti-dyons
pair into neutral ``dyon-antidyon molecules" bound through fermion exchanges~\cite{SHURYAK}, 
a situation somewhat reminiscent of the BKT transition~\cite{ZHITNITSKY,FATEEV}.

This can be seen by noting that the chiral matrix $\tilde{\bf T}$ in (\ref{T12}) is banded with
a band range set by the inverse temperature

\be
{\bf T}_{ij}\approx {\bf t}_f\,e^{-\frac 12 \omega_0\,|x_i-y_j|}\rightarrow{\bf t}_f \,\delta_{ij}
\label{RANGE}
\ee
With increasing temperature, the range  of $\tilde{\bf T}$ is reduced to the nearest neighbor. 
As a result, the hopping is stalled with

\be
\left|{\rm det}\,\tilde{\bf T}\right|\rightarrow \left|{\bf t}_f\right|^{K_L+K_{\bar L}}\,\delta_{K_LK_{\bar L}}
\nonumber
\ee
The light quark spectrum is now gapped at $\lambda_\pm=\pm |{\bf t}_f|$ with a vanishing chiral condensate
$\left<\bar q q\right>=0$. An estimate of the hopping parameter follows from 

\be
{\bf t}_f\equiv  \int\frac{d^3p}{(2\pi)^3}\,{\bf T}(p) \approx 0.8\,\omega_0
\label{TF}
\ee
using (\ref{TIJ}-\ref{DZERO}) and Parseval equality.

A simple but crude estimate of the transition density  at which the pairing into
molecules overtake the chirally broken phase is when the molecular gap becomes larger 
than $\Sigma$, thus restoring chiral symmetry. $\Sigma$ characterizes the
size of the delocalized zero mode zone. Using (\ref{TF}) this occurs for
$ |{\bf t}_f|\approx 2.51T\approx \Sigma$. Since
$\Sigma/T=n_D/|\left<\bar q q\right>|$, this means a transition when 
$|\left<\bar q q\right>|/n_D\approx 1/2.5$. From the numerical fit (\ref{QFIT}) this 
estimate yields to a chiral restoration for 
a  dilute dyon ensemble with
\be 
{n_D \over T^3} < {n_\chi \over T^3}\approx 0.16 
\ee
Near the transition temperature,  a substantial amount of dyons can already be paired, 
 resulting  in a  weakening of the chiral condensate.

In terms of (\ref{RANGE}) the partition function (\ref{SU2}) is highly correlated. The
result after summing over pairs is

\bea
{\cal Z}_{\rm mol}[T]=&&\int D[b]\, D[\sigma]\, D[\chi]\, D[w]\, e^{-S_0-S_M}\nonumber\\
&&\times\,|{\bf t}_f|\,\sqrt{F_LF_{\bar L}}\,I_2(|{\bf t}_f|\sqrt{F_LF_{\bar L}})
\label{MOL}
\eea
with $S_0$  defined in  (\ref{FREE1}-\ref{FREE2}) and $S_M$ defined in  (\ref{FREE3}) for only
$M$ and $\bar M$. The argument of the modified Bessel function ${I}_2$  is composed of

\bea
F_L=&&\int d^3x\,  e^{+b-i\sigma+w_L-w_M}\nonumber\\
&&\times f_L\left(4\pi v_l+|\chi_L    -\chi_M|^2+v_L-v_M\right)\nonumber\\
F_{\bar L}=&&\int d^3x\,  e^{+b+i\sigma+w_{\bar L}-w_{\bar M}}\nonumber\\
&&\times f_{\bar L}\left(4\pi v_{\bar l}+|\chi_{\bar L}    -\chi_{\bar M}|^2+v_{\bar L}-v_{\bar M}\right)\nonumber\\
\label{MOLX}
\ee

The molecular partition function in (\ref{MOL}) is highly non-linear in the auxiliary fields. 
For large $|{\bf t}_f|$, we may use the asymptotic form of $I_2(z)\approx e^{z}/\sqrt{2\pi z}$ in (\ref{MOLX}) and linearize the
argument of the modified Bessel function

\be
\sqrt{F_LF_{\bar L}}\approx \frac{F_L+F_{\bar L}}{\sqrt{2}}\left(1-\frac{F_L^2+F_{\bar L}^2}{2\,(F_L+F_{\bar L})^2}\right)
\label{LIN}
\ee
As a first step to be justified below, we may drop the non-linear contributions  in (\ref{LIN}) and the
pre-exponent in (\ref{MOL}) to have

\bea
{\cal Z}_{\rm mol}[T]\approx \int D[b]\, D[\sigma]\, D[\chi]\, D[w]\, e^{-S_0-S_M+({\tilde{F}_L+\tilde{F}_{\bar L}})/\sqrt{2}}\nonumber\\
\label{MOLXX}
\eea
after rescaling $f_{L,\bar L}\rightarrow f_{L,\bar L}\,|{\bf t}_f|/\sqrt{2}$ in $\tilde{F}$. (\ref{MOLXX}) is now analogous to the SU(2)
Yang-Mills partition function~\cite{DP,LIU1} with the new re-scaled fugacities. A rerun of the arguments in this case shows that
the ground state is still center symmetric. However, the ground state is chirally symmetric. It is worth noting that
this state is parity even, so the neglected non-linear corrections in (\ref{LIN}) amounts to $(1-1/4)$ which is about a 25\% reduction
in the pertinent pressure contribution which is then

\be
{\cal P}_{\rm mol}=\frac{{\rm ln}{\cal Z}_{\rm mol}}{V_3/T}\approx 8\pi T\left(\frac{f_Mf_L\,|{\bf t}_f|}{\sqrt{2}}\right)^{1/2}
\label{MOLPRESSURE}
\ee

\section{Higher number of colors and flavors}

The extension of the current analysis to many $N_c$ colors and $N_f$ massless flavors
is straightforward in principle. For finite $N_c$  the KvBLL instanton
splits into $N_c$ constituent dyon with $1/N_c$ topological charge and fugacity $f_l$ with $1\leq l\leq N_c$.
The L-dyon zero mode which is anti-periodic is now carried by the $l=N_c$ constituent dyon.
The net effect is a change in the fermionic contribution in (\ref{NEWS})
through ${\bf G}\rightarrow {\bf G}\otimes {\bf 1}_f$, and a change in the parity even
effective potential (\ref{POT}) as

\bea
-{\cal V}/\mathbb{V}_3\rightarrow &&+i\lambda N_f\Sigma
+4\pi  f_iv_i\,( e^{w_{i+1}-w_{i}}+e^{w_{\bar {i}+1}-w_{\bar i}} ) \nonumber \\
&&+4\pi f_{N_c}v_{N_c}\,\frac 1{N_f!}{\rm det}_{N_f}(\psi^\dagger_l \gamma_+ \psi_g)\,e^{w_{N_c}-w_{N_c+1}}\nonumber\\
&&+4\pi f_{\bar {N_c}}v_{\bar {N_c}}\,\frac 1{N_f!}{\rm det}_{N_f}(\psi^\dagger_l \gamma_- \psi_g)\,
e^{w_{\bar {N_c}}-w_{\bar {N_c}+1}}\nonumber\\
\label{DETER}
\eea
where the implicit  i-summation is over $i=1,...,N_c-1$. The
potential ${\cal V}$ has manifest ${\rm SU}(N_f)_V\times {\rm SU}(N_f)_A\times {\rm U}(1)_{V}$
flavor symmetry. 
As a result, the parity even effective potential (\ref{SU2POT1}) after pertinent bosonization
and Fierzing yields

\bea
\label{SU2POT10}
- {\cal V}/\mathbb{V}_3\rightarrow&& +i\lambda N_f\Sigma +2\alpha(N_c)\,\Sigma^x
\\&&+N_f \int \frac{d^3p}{(2\pi)^3}{\rm ln}\,\left(1-\lambda^2{{\bf T}^2(p)}\right)\nonumber
\eea
with $x=N_f/N_c$ and $\alpha(N_c)=4\pi f(N_c)/2^x$.
The mean fugacity is

\be
f(N_c)=\left(f_1....f_{N_c}\right)^{1/N_c}
\ee
We note that its scaling with $N_c$ follows from the scaling of each fugacity 
by semi-classics, i.e. $f_i\approx 1/\alpha_s^2$. Thus

\be
f(N_c)\approx N_c^{2N_c/N_c}\approx N_c^{2}
\ee
so  that $\alpha(N_c)\approx N_c^2$.

\subsection{Gap equation and chiral condensate}

For general $x=N_f/N_c$,   the saddle point equation in $\Sigma$ of (\ref{SU2POT10}) gives

\be
\Sigma=\left(\frac{\tilde\lambda}{2x\alpha(N_c)}\right)^{\frac 1{x-1}}
\label{SX}
\ee
after the shift $-i\lambda\rightarrow\lambda$ and $\tilde\lambda=N_f\lambda$.  With this in mind and inserting (\ref{SX}) into (\ref{SU2POT10})
yields 

\bea
\label{XVX}
-{\cal V}/\mathbb{V}_3=&&-2\alpha(N_c)\,(x-1)\left(\frac{\tilde\lambda}{2x\alpha (N_c)}\right)^{\frac{x}{x-1}}\\
&&+xN_c\int \frac{d^3p}{(2\pi)^3}{\rm ln}\left(1+\frac{\tilde\lambda^2}{N_f^2}{\bf T}^2(p)\right)\nonumber
\eea
The case $x=1$ is special. The effective potential in (\ref{SU2POT10}) is linear in $\Sigma$ with no a priori saddle point
along $\Sigma$. We have checked that taking the saddle point in $\tilde\lambda$ first, and then the saddle-point in $\Sigma$ after the
substitution results in the same gap equation to follow. Also, it can be checked explicitly that the same results follow
by taking the limit $x\rightarrow 1$.

The effective potential (\ref{XVX}) has different shapes depending on the ratio of the number of flavors to the number  of  
colors $x$. Let us explain that in details for four cases:

(i) If $x<1$ the first term in (\ref{XVX}) has a positive coefficient and a negative power,
so it is decreasing at small $ \tilde\lambda$. At large value of $ \tilde\lambda$
the second term is growing as ${\rm ln}\tilde\lambda$. Thus a minimum in between 
must exist. This minimum is the physical solution we are after.

(ii) If  $1<x<2$ the coefficient of the first term is negative but its power is now positive.
So again there is a decrease at small $ \tilde\lambda$ and thus a minimum.

(iii) If $x>2$  the leading behavior at small 
$ \tilde\lambda$ is now dominated by the second term which goes as $ \tilde\lambda^2$
with positive coefficient. One may check that the potential is
monotonously increasing for any $\tilde\lambda$ with  no extremum. 
There is no gap equation, which means  chiral symmetry cannot be broken
in the mean-field approximation.

(iv)
If $x=2$ there are two different contributions of opposite sign to order $\tilde\lambda^2$ at small
$\tilde\lambda$. An extremum  forms only if the following condition is met

\be
\int\frac{d^3p}{(2\pi)^3}{\bf T}^2(p) \,< \frac {N_c}{4\alpha(N_c)}={\cal O}\left(\frac 1{N_c}\right)
\label{XE2}
\ee
Using the exact form (\ref{Anp}) and the solution to the gap equation  at $T=T_0$, we have

\be
\int\frac{d^3p}{(2\pi)^3}{\bf T}^2(p) = \frac{10.37}{T_0}
\ee
which shows that (\ref{XE2}) is in general upset, and this case does $not$ possess a minimum. 

With this in mind and for $x<2$, the extremum of (\ref{XVX}) in $\tilde\lambda$ yields the gap-like equation

\be
\left(\frac{\tilde\lambda}{2x\alpha (N_c) }\right)^{\frac{x}{x-1}}=
\frac{N_cV_{N_f-1}}{\alpha(N_c)}
\label{NVC}
\ee
with the new identification

\be
V_{N_f-1}\equiv \int \frac{d^3p}{(2\pi)^3}
\frac{\frac{\tilde\lambda^2}{N_f^2}{\bf T}^2(p)}{1+\frac{\tilde\lambda^2}{N_f^2}{\bf T}^2(p)}
\label{VNF}
\ee
The dyonic density is now identified with 

\be
n_D=2N_c\frac 12 \frac{\partial ({-\cal V}/{\mathbb V_3})}{\partial {\rm ln} f_M}=2N_cV_{N_f-1}
\label{NDF}
\ee
In terms of (\ref{VNF}-\ref{NDF}) the running constituent mass $M(p)=(\tilde\lambda/N_f)\,p{\bf T}(p)$
obeys the gap equation

\be
\int\frac{d^3p}{(2\pi)^3}\frac{M^2(p)}{p^2+M^2(p)}=\frac {n_D}{2N_c}
\label{GAPNCNF}
\ee
From (\ref{SX}) it follows that $\tilde\lambda\approx N_cN_f$. 
From (\ref{NVC}) it follows that $V_{N_f-1}\approx N_c$ and therefore $n_D\approx N_c^2$. The dyonic description we have reached is consistent with large $N_c$ counting. Since $n_D\approx N_c^2\gg 1$ crystallization in the form of dyonic salt is expected at large $N_c$~\cite{SIN}.


For $x< 2$ the center symmetric vacuum also breaks spontaneously chiral symmetry, with a vacuum
condensate given by

\be
\frac{\left<\bar q q\right>}T=-2\tilde\lambda_0
\label{QQ}
\ee
$\tilde\lambda_0$ is the value of $\tilde\lambda$ in the chiral limit.
Since $\tilde\lambda_0\approx N_fN_c$ the chiral condensate in (\ref{QQ}) is of order $N_fN_c$ as expected.

To summarize: Chiral restoration as well as the loss of center symmetry
occur  simultaneously for $x_\chi\geq 2$ as per our result in (\ref{SU2POT}). 
This value of $x=N_f/N_c$ is  close to the the critical value of $x_\chi=5/3$ originally suggested in the instanton liquid model
\cite{ALL} (first reference). First simulations of the dyon ensemble with $N_c=2$~\cite{Faccioli:2013ja} also
indicate that the border line seems to be $N_f=4$, in agreement with $x_\chi\approx 2$.

Current lattice data are summarized e.g. in Fig.5 of \cite{Lombardo:2014fea} for $N_c=3$.
Indeed, they seem to  indicate a change in the value of the chiral transition temperature
(in units of the vacuum string tension) $T_\chi/\sqrt{\sigma}$ at $x=2$ or $N_f=6$,
but instead of vanishing, this ratio remains flat up to $N_f=8$. For such a large $N_f$
the number of quark lines $2N_f$ connected to an $L$ dyon is  large. Maybe 
the correlations between them are too strong for the mean field approximation to remain valid.   

The chiral transition we have discussed in this section should not be confused with another phase transition in theories with
a large number of flavors, namely the conformal (fixed infrared coupling) phase. 
The  reported lattice~\cite{LAT} and holographic  (Veneziano limit) \cite{HOLO} results put this 
conformal transition at
a much larger number of flavors $x_{\rm conformal}\approx 4$, or $N_f=12$ for $N_c=3$.

\subsection{Thermodynamic of dyonic phase with $x\leq 1$}

\begin{figure}[h!]
  \begin{center}
  \includegraphics[width=7cm]{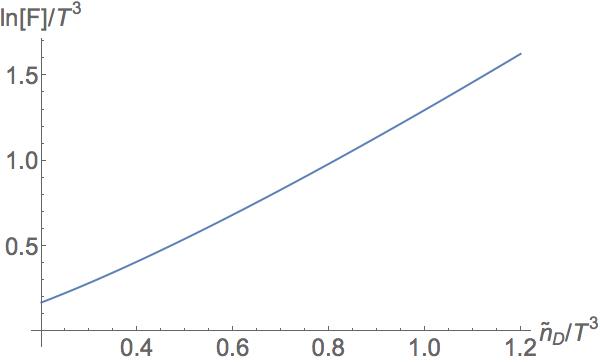}
   \caption{The fermionic loop ${\rm ln}[F]/T^3$ versus the reduced dyon density $\tilde{n}_D/T^3$.  See text.}
     \label{fig_ferm}
  \end{center}
\end{figure}

In the presence of light quarks, the total thermodynamical pressure of the dyon-anti-dyon liquid  consists of the classical and non-perturbative
contributions in  (\ref{SU2POT10}) at the extremum, plus its perturbative correction for finite
and symmetric holonomies $v=1/N_c$~\cite{WEISS}, plus the purely perturbative
black-body contribution (ignoring the higher order ${\cal O}(\alpha_s)$ quantum corrections). 
Identifying  the classical pressure with $-{\cal V}/\beta V_3$ with $\beta=1/T$, we have

\bea
\label{PCLA}
\frac{{\cal P}_{\rm tot}-{\cal P}_{\rm per}}{N_cT^4}=&&+(1-x)\,\tilde{n}_D\\&&
+\frac{x}{T^3}\int \frac{d^3p}{(2\pi)^3}{\rm ln}\,\left(1+\frac {M^2(p)}{p^2}\right)\nonumber
\eea
with $\tilde{n}_D=n_D/(N_cT^3)$.
The assessment of the logarithmic integral follows by numerical integration 
using the explicit form of $M(p)$ and the solution to the gap equation.  The result
is linear in the reduced dyon density for small and asymptotic densities

\bea
\frac 1{T^3}\int \frac{d^3p}{(2\pi)^3}\,{\rm ln}\left(1+\frac {M^2(p)}{p^2}\right)\approx \kappa(\tilde{n}_D)\,\tilde{n}_D
\label{LOG}
\eea
with $\kappa (\tilde{n}_D\ll 1)\approx 1$ and $\kappa(\tilde{n}_D\gg 1)\approx 2$. A simple
interpolation  to the overall numerical results is

\be
\kappa(\tilde{n}_D)\approx \frac{1+2\frac{\tilde{n}_D}{10}}{1+\frac{\tilde{n}_D}{10}}
\ee
Since $\tilde{n}_D\approx {\cal O}(N_c)$, large density corresponds to large
$N_c$ with $\kappa\approx 2$, modulo crystalization.
In  Fig.~\ref{fig_ferm} we display (\ref{LOG}) as a function of the reduced dyon density $\tilde{n}_D$ at
 intermediate densities.
The linearity of the logarithm in $\tilde{n}_D$ both at small and asymptotic dyon densities follow from the scaling
of $V_0=n_D/2N_c$ with $\lambda$ as discussed in (\ref{V0}).

In terms of (\ref{LOG}) the classical pressure contribution in (\ref{PCLA})  simplifies to

\bea
\label{PCLAX}
\frac {{{\cal P}_{\rm tot}-{\cal P}_{\rm per}}}{N_cT^4}\approx (1-x(1-\kappa ))\, {\tilde n}_D
\eea
with $1\leq \kappa\leq 2$.
In the quenched limit or $x=0$ it reduces to the dyonic result obtained for the pure Yang-Mills analysis
in~\cite{LIU1} ignoring the Debye-Huckel corrections. The fermion induced interactions in the center
symmetric phase increase  the pressure away from the free limit for $0<x\leq 1$. Remarkably, 
for small densities$(1-\kappa)\approx 0$ and (\ref{PCLAX}) with fermions is close to a free ensemble
of dyons. For large densities or large $N_c$, $(1-\kappa)\approx -1$ and (\ref{PCLAX}) with  fermions is more repulsive
than the free dyon ensemble. 

The perturbative contribution is given by

\bea
\frac{{\cal P}_{\rm per}}{T^4}&&\approx
-\frac{\pi^2}{45}\left(N_c^2-\frac 1{N_c^2}\right)+\frac{\pi^2}{45}\left(N_c^2-1\right)\nonumber\\
&&-\frac{7\pi^2x}{180}\left(N_c^2-\frac 1{N_c^2}\right) + \frac{7\pi^2x}{180}N^2_c
\label{EOS2}
\eea
The first  contribution is the free gluon contribution in the symmetric
phase with $v=1/N_c$. The second  contribution is the free black-body gluon contribution,
which is cancelled by the second contribution in leading order in $1/N_c$ in the
symmetric phase~\cite{DP}. The third contribution is the free quark contribution in the
symmetric phase with $v=1/N_c$. The fourth and last contribution is the black-body
quark contribution, which we note is cancelled by the third contribution in leading order in
$1/N_c$. This generalizes the observation in~\cite{DP} to QCD.

An estimate of the transition temperature $T_c$
from the symmetric phase with $v=1/N_c$ to the asymmetric phase  with $v=0$
follows  when all the non-black-body contributions in the total pressure ${\cal P}_{\rm tot}$ cancel out.  This
occurs when  the  rescaled dyon density ${\bf n}_D=n_D/(N^2_cT^3)$ solves

\be
 {\bf n}_{Dc}\approx 
 \frac{\pi^2}{45}\left(1-\frac 1{N_c^4}\right)\frac{1+\frac {7x}4}{1+x(\kappa({\bf n}_{Dc})-1)}\nonumber\\
\label{NCC}
\ee
with again $1\leq \kappa({\bf n}_{Dc})\leq 1$.

\subsection{Thermodynamic of  molecular phase}

For completeness we note that near the chiral transition most of the $L\bar L$ dyons start
to pair into molecules, for which case the total pressure is more appropriately described by

\bea
\frac{{\cal P}_{\rm tot, mol}-{\cal P}_{\rm per}}{T^4}\approx \frac{2{\tilde \Lambda}^4}{\alpha_s^2} 
\left(\frac{|\tilde{\bf t}_f|^{N_f}}{\sqrt{2}}\right)^{\frac 12}
\label{EOS2M}
\eea
with $|\tilde{\bf t}_f|=|{\bf t}_f|/\Lambda$. Here, the scale parameter 
$\tilde\Lambda=\Lambda/T$ is identified with the vacuum dyon density $n_D\rightarrow 2\Lambda^4/\alpha_s^2T$
as in~\cite{DP}. We note that for $N_f\rightarrow 0$, (\ref{EOS2M})
is off by ${2}^{-\frac 14}$ in the ground state pressure from the Yang-Mills limit in~\cite{DP}. This
can be traced back to our linearized approximation in (\ref{LIN}).
The critical temperature is now

\be
T_c\approx \frac{2\Lambda}{\sqrt{ \tilde \alpha_s}}
\left(\frac{|\tilde{\bf t}_f|^{N_f}}{\sqrt{2}}\right)^{\frac 18}\,\frac 1{{h^{\frac 14}(x)}}
\label{EOS4M}
\ee
with $\tilde\alpha_s=N_c\alpha_s$.

\bea
h(x)=\frac{\pi^2}{45}\left(1+\frac{7x}{4}\right)\left(1-\frac 1{N_c^4}\right)
\eea
(\ref{EOS4M}) characterizes the transition from a center symmetric but chirally symmetric phase to a center
asymmetric and chirally symmetric phase. Which likely transition is to occur first can be estimated by comparing
the total liquid pressure in (\ref{PCLA}) to the total molecular pressure in (\ref{EOS2M}). This is best addressed 
using mixtures.


\section{Conclusions}

We have extended the mean field treatment of the SU(2) dyon-anti-dyon liquid in~\cite{LIU1},
 to account for light quarks. Anti-periodic fundamental quarks  develop zero modes for 
the $ {\rm L},\bar {\rm L}$-dyons only.  In the dense  phase under consideration with $T<T_c$,
these zero modes are collectivized into a Zero Mode Zone of quasi-zero modes which dominates the
low-eigenvalue part of the Dirac spectrum. This phenomenon is analogous to the one used in the instanton liquid model~\cite{ALL}, 
although the zero modes themselves and most of the results are different. The important interplay between center symmetry
and the spontaneous breaking of chiral symmetry which is absent in~\cite{ALL}
is now clarified. In particular, we have explicitly shown how the chiral effective 
Lagrangian for light quarks knows about confinement.

In the infrared, the fermionic determinant
is entirely saturated by these quasi-zero modes~\cite{SHURYAK}, modifying the dyon-anti-dyon measure initially suggested
in~\cite{DP,DPX} to include light quarks. 
For the $N_f=1$ case of one massless quark, we have shown that the fermionic determinant
 modifies the  L,$\bar {\rm L}$-dyonic fugacities through chiral  fermionic bilinears  that upset the U$_A(1)$
symmetry. By a series of bosonic and fermionic techniques we have explicitly mapped the interacting
dyon-anti-dyon Coulomb liquid with light quarks on a 3-dimensional effective theory with fermions. The
translationally and parity invariant ground state was shown to follow from pertinent gap equations.
The ground state breaks spontaneously chiral symmetry by developing a fermion condensate. For $N_f=1$  it does
not produce a Goldstone mode because of the U$_A(1)$ anomaly. We have derived explicit expressions and estimates for masses
of the $\sigma$ and $\eta$ mesons.

We have shown how the model generalizes to arbitrary number of flavors and colors. 
In the whole temperature interval in which our approach is applicable, 
the ensemble is center symmetric (confining) and  breaks spontaneously
chiral symmetry  provided $x=N_f/N_c<x_\chi\approx 2$. The loss of center symmetry and chiral symmetry restoration 
in this model seem to occur
simultaneously for $x\geq 2$. We have noted that in the case of a very large $N_f$ the fermion-induced interactions
maybe too strong to trust the mean field approximation we used. This point needs to be pursued numerically on the lattice.

This conclusion can be compared to the critical value of $x_\chi\approx 5/3$ of
the numerical simulation of the instanton liquid model~\cite{ALL}( first reference). 
The first simulation of the dyon ensemble with fermions, for $N_c=2$ \cite{Faccioli:2013ja}, found
 the border line case is $N_f=4$, also in agreement with $x_\chi\approx 2$. So whether the
 transition is an artifact of the mean field approximation or not remains to be studied.

The chiral transition should not be confused with the transition to conformal -- fixed infrared coupling-- phase,
for which current  lattice and holographic results put this transition at a
 much larger number of flavors $x_{\rm conformal}\approx 4$, or $N_f=12$ for $N_c=3$.

Near and above the chiral transition  the fermionic correlations are strong enough
to pair L-dyons with $\bar{\rm L}$-anti-dyons into molecules. The dilute regime  involved
 has been explored  numerically in~\cite{SHURYAK}. In this paper we only produced 
 some estimates of the transition parameters.
 
In our approach the confining 
and  chirally broken phase have been treated via the mean field approximation only, 
so the resulting gap equation has either a finite or zero $\Sigma$, with a finite jump.
In the future one can probably  include  the ${\rm L\bar L}$  correlations in the ensemble.
 The result would be a depletion of  the chiral condensate
with perhaps a more continuous cross-over transition as currently  observed in 
QCD-like theories with several flavors of massive quarks.

In the extreme case where all L-dyons and $\bar{\rm L}$-anti-dyons 
pair to molecules, we have shown that the linearized molecular partition function 
supports a phase with center symmetry but restored chiral symmetry.
It would be interesting in the future to see if such a phase may exist, at some $N_c,N_f$.
At this moment, lattice data on that issue are also not clear, see e.g. \cite{LAT}.

An estimate of the pressure in the center symmetric phase shows that both the free gluon and fermion
loop nearly cancel out in leading order in $N_c$. We
have used it to estimate the transition density  from a center symmetric  phase to a
phase with broken center symmetry. A similar estimate of the transition density
was made in~\cite{LIU1} in the absence of fermions.


The current model can be expanded and improved in a number of ways.
The current analysis has been done for the sector
with zero $\theta$. As indicated earlier, the L-zero modes were selected
over the M-zero modes, creating a topological unbalance and a
lack of manifest $\theta$-periodicity in the induced effective action. This  can be generalized to arbitrary $\theta$ angle 
through an  extended formalism. 

Also, we have not included here some Coulomb corrections discussed in~\cite{LIU1}, to keep the
analysis simpler and  to illustrate the interdependence of the center symmetry  and the chiral symmetry breakings
in this model. These corrections are Debye-like and still within the semi-classical analysis.

Some improvement of the moduli space metric may be considered in the future.
We recall that the moduli space metric used in (\ref{SU2})  while exact for $\rm LM$ dyons at all
separations, is only exact asymptotically for $\rm LL,MM$ dyons. While the formers attract, the latters repel. In
the center symmetric or confining phase, we expect the like-dyons to stay away from each other while
the un-like dyons  to mingle and screen.  
In the dyon-antidyon channels the treatment is so far classical only, with one-loop
effects absent.
We hope to report on some of these issues, as well as on a full analysis of 
the meson spectrum for the dyon-anti-dyon liquid with $N_f>1$ next.

\section{Acknowledgements}

This work was supported by the U.S. Department of Energy under Contracts No.
DE-FG-88ER40388.

 \vfil

\end{document}